# DESIGN OF A DISTRIBUTED CONTROL SYSTEM BASED ON CORBA AND JAVA FOR A NEW RIB FACILITY AT LNL

Stefania Canella, Giorgio Bassato, L.N.L. - I.N.F.N. Legnaro (Pd), Italy
canella@lnl.infn.it, bassato@lnl.infn.it

Abstract

SPES (Study for the Production of Exotic Species) [1] is a L.N.L. project that will produce by the end of this year the design of a facility for Radioactive Ion Beams (RIBs) originated by fission fragments produced by secondary neutrons; it will be characterized by moderate size, performance and cost and will produce also intense neutron beams for activities both in fundamental and applied Nuclear Physics.

In the context of this design study and tightly related to the medium size of this facility, the architecture of a distributed control system using the Common Object Request Broker Architecture (CORBA, [2]) as middleware framework and Java as main programming language was investigated for the core components (diagnostics, optics, RF-control) of the primary accelerator.

The performances of CORBA middleware for the high level control system were measured in different conditions and showed to be sufficient to cover the requirements for remote operations (all feedback loops will be performed either in specialized hardware or by dedicated real-time embedded controllers).

A minimal programming effort, a good level of modularity and long-term maintenance were some of the reasons to choose Java [3] and its related Integrated Development Environments (IDEs) as the main programming language and as a software platform for the Graphical User Interface (GUI) and the middleware implementations of this project.

## 1 SPES PRIMARY ACCELERATOR

The primary accelerator of SPES is a LINAC for a high intensity proton beam (1-30 mA) up to an energy of 100 MeV. Fission fragments (RIBs) produced by the flux of neutrons will be then boosted by RFQs (Radio Frequency Quadrupole) structures and injected in an L.N.L. LINAC operating since 1995.

The base components of this primary accelerator:
- a proton source (up to 80 mA) at 80 KV;
- a normal conductive RFQ at 5 MeV as low energy booster for a 30 mA proton beam;
- an ISCL (Independently phased Superconducting Cavity Linac) up to the final energy of 100MeV.

The primary accelerator detailed design and construction will be performed in 2 steps: the first up to an energy of 10 MeV and the second upgrading the final energy from 10 to 100 MeV.

By the end of this year a final approval for the first step is expected.

## 2 THE CONTROL SYSTEM OF THE PRIMARY ACCELERATOR

### 2.1 General Layout

The core components (beam diagnostics, beam optics, RF-control) of the primary accelerator will be controlled by a distributed system, that is by a set of heterogeneous computers, modular systems, embedded controllers and field devices whose *objects* (data, commands, trends, alarms ...) have to be sharable among local operators and (with some restrictions, to be defined) remote users.

### 2.2 Hardware architecture

From the hardware (HW) point of view, a standard three-level architecture is foreseen.

PCs and/or Workstations (WS), at the top level, will be used through graphic interfaces by local operators and remote users, for data storage and general services such as network routing and firewall, on-line documentation and printing.

At the middle level there will be VME based systems with PowerPC processor boards and any other specialized board suitable to manage IN/OUT analog and digital signals directly connected to the field.

Embedded controllers and field devices at the bottom level will perform local self-contained tight interactive activities and fast feedbacks.

The connection between top and middle level is planned through a standard fast ethernet (switched 100BaseT), while different heterogeneous connections using different protocols (thin wire ethernet, fieldbuses, serial lines) will be used between middle level systems and field devices or embedded controllers.

## 2.3 Software architecture

Different operating systems (Solaris, Linux, Windows) will be equivalent environments for the top level systems, while VxWorks will be used as the main real-time operating system for the middle level layer and for embedded controllers.

Java applications and applets on PCs and WSs (Java platforms) will establish the SW framework for the GUI and general services. C or C++ will be the basic programming languages for applications for real-time VME systems and for embedded controllers.

Communications on the network will be based on CORBA middleware. Low level transport mechanisms such as BSD sockets will be also used, if necessary.

## 2.4 Reasons for CORBA and Java IDL

There are several reasons to use CORBA:
- CORBA is a standard developed since 1989 to operate across different networks and operating systems;
- CORBA objects may be located anywhere on a network and may communicate each other no matter where they are located, this makes it an ideal framework for distributed applications;
- CORBA components may be supplied by different vendors, as a standard Interface Definition Language (IDL) has to be used to define the interfaces to CORBA objects;
- CORBA objects may be written in different languages (Java, C++, C).

Some good reasons lead also to use Java IDL, at least for top level applications:
- Java IDL is an ORB (Object Request Broker) provided with the Java 2 Platform: it can be directly used to define, implement and access CORBA objects from the Java programming language;
- together with Java IDL, a Transient Name Server (tnameserv) is provided on any Java2 Platform.

Java IDL ORB is then a simple tool to test CORBA [4] middleware prototypes and their performances.

## 2.5 Reasons for Java

There are many reasons to use Java, where possible, in place of traditional procedural languages:
- Java is object-oriented and designed for network based distributed software;
- Java bytecode is directly usable on any Java platform (Java *compiled* applications are really portable across multiple HW-SW platforms);
- Java is multithreaded, that is suited for applications performing multiple concurrent activities;
- Java is an ideal language for CORBA programming as CORBA objects may be immediately created and used on a Java 2 Platforms.

Besides all this, Java includes many libraries of objects providing extended functionalities on I/O, network interfaces, data base access, graphics.

## 2.6 IDEs for Java Platforms

Today there are many IDEs available for medium-large software projects on Java Platforms, especially to produce sophisticated GUIs.

The most promising IDE for CORBA-Java distributed applications is probably JB4 (JBuilder 4) from Borland and Forte from SUN. JB4 and Visibroker ORB are currently under test for SPES. Forte for JAVA will be also tested in the next months.

## 2.7 Performances of CORBA on Java Platforms

In order to test CORBA middleware and Java IDL environment some Java tasks were implemented, the most interesting of them being a client-server couple including the capability to measure the elapsed time of a typical control transaction: the trip-time of a message between CORBA-based Java applications, eventually on the network (the trip being the transit of a message from the client to a server and back). These special couple of tasks was developed to measure the typical delay that has to be expected for a service request in a CORBA-based distributed control system. This application includes an IDL module describing the available interface, a transient server (a program that contains the implementations of the IDL interface), and the code of a CORBA client invoking the available operations on distributed objects. As naming service the standard *tnameserv* utility was used. The length of the message was changed in a wide range of meaningful values. This test tool was used in four different HW-SW configurations:
- Test n. 2: all tasks (nameserver, server and client) running on the same Sun-Solaris8 ULTRA10 WS;
- Test n. 3: server and nameserver running on a Sun-Solaris8 ULTRA10 WS, client running on another Sun-Solaris8 ULTRA10 WS, the two WS connected by a 10 Mb/s ethernet link with no other traffic;
- Test n. 4: server, nameserver and client running on a PC-linux (with an old 200MHz processor);

- Test n. 5: server and nameserver running on a PC-linux (200MHz processor), client running on Sun-Solaris8 ULTRA10 WS, the two computers connected by a 10 Mb/s ethernet link with no other traffic.

All tests were performed by a client sending a message of fixed length and receiving its echo from the server: the trip-time was taken by the client as the elapsed time between the *send* operation and the *receive* operation. Each test was repeated 500 times to have a reasonable statistics. The results (mean, minimum and maximum elapsed times in ms) are summarized in Table 1.

## 3 SUMMARY

The performances of CORBA middleware in the above described tests proved to be sufficient to cover the requirements for remote operations: a typical message of 1 Kbyte (a request with some parameters) will travel forward from the client to a server and back (with the answer to the request of service) in an average time not greater than 10 ms, in all the four configurations. A greater delay may be noticed from time to time due to the concurrent activities of the systems, but always above the acceptable limit of 100 ms.

From the above described basic choices (CORBA, Java, IDEs) a number of benefits are expected for the control system of SPES primary accelerator:
1. a long lifetime (over 10 years) both from the HW and SW point of view;
2. HW components at the top and middle levels will be available from a wide range of manufacturers;
3. high SW reliability because compile-time and run-time extensive checking will help to have a fast and extensive code debugging;
4. Java automatic memory garbage collection will also help to produce reliable code;
5. minimum time to repair (HW) and/or for maintenance (HW/SW)
6. uniform code (and bytecode) for different Java platforms greatly simplify SW updating and maintenance.

|  | Test n. 2 (1 WS Sun) | | | Test n. 3 (2 WS Sun + net) | | |
|---|---|---|---|---|---|---|
| Msg length | Mean [ms] | Min [ms] | Max [ms] | Mean [ms] | Min [ms] | Max [ms] |
| 1 byte | 2.58 | 1 | 27 | 2.18 | 1 | 13 |
| 10 byte | 2.19 | 1 | 42 | 2.03 | 1 | 23 |
| 100 byte | 2.73 | 1 | 151 | 2.54 | 1 | 97 |
| 1K byte | 1.89 | 1 | 17 | 5.57 | 5 | 12 |
| 10K byte | 5.95 | 4 | 60 | 25.67 | 23 | 124 |
| 100K byte | 58.2 | 41 | 133 | 255.69 | 219 | 400 |
| 1M byte | 575.34 | 553 | 1127 | 2726.44 | 2553 | 3363 |
|  | Test n. 4 (1 PC linux) | | | Test n. 5 (1 WS + 1 PC + net) | | |
| Msg length | Mean [ms] | Min [ms] | Max [ms] | Mean [ms] | Min [ms] | Max [ms] |
| 1 byte | 4.73 | 3 | 55 | 3.35 | 2 | 57 |
| 10 byte | 4.12 | 3 | 93 | 3.32 | 2 | 46 |
| 100 byte | 4.81 | 3 | 313 | 3.98 | 3 | 99 |
| 1K byte | 4.74 | 4 | 40 | 8.15 | 7 | 14 |
| 10K byte | 22.09 | 19 | 244 | 49.9 | 47 | 74 |
| 100K byte | 487.46 | 448 | 752 | 781.75 | 495 | 910 |
| 1M byte | 2710.68 | 2646 | 5651 | 6150.69 | 5763 | 6953 |

Table 1 - Test results (trip-time for messages of different lengths) for CORBA-Java based applications in 4 different configurations